\documentclass[reprint,superscriptaddress,amsmath,amssymb,aps]{revtex4-2}

\usepackage{booktabs}
\usepackage{dcolumn}% Align table columns on decimal point
\usepackage{tikz}
\usepackage{supertabular}
\usepackage{changepage}
\usepackage{siunitx}
\usepackage[colorlinks=true, 
    colorlinks=true,
    linkcolor=blue,
    filecolor=magenta, 
    citecolor=blue,
    urlcolor=black]{hyperref}
\usepackage[version=4]{mhchem}
\usepackage{float}
\usepackage{caption}
\usepackage{subcaption}
\usepackage{xfrac}
\usepackage{xr-hyper}
\usepackage{hyperref}
\usepackage[version=4]{mhchem}

\begin{document}

\title{Can Strain or Anion Interchange Make an Unstable Structure Stable? Energetics, Lattice Dynamics and Strain-Tunable Band Gaps of Lithium Chalcohalide Antiperovskites (Li$_{3}$$BA$) and their Anion Interchange Variants (Li$_{3}$$AB$)} 
\author{Ismail A. Buliyaminu}
\affiliation{Department of Physics \& Astronomy, Michigan State University, East Lansing, Michigan 48824, United States.}
\author{Ehsan Gowdini}
\affiliation{Department of Physics \& Astronomy, Michigan State University, East Lansing, Michigan 48824, United States.}
\author{Phillip Duxbury}
\affiliation{Department of Physics \& Astronomy, Michigan State University, East Lansing, Michigan 48824, United States.}
\author{Jose L. Mendoza-Cortes}
\email{jmendoza@msu.edu}
\affiliation{Department of Physics \& Astronomy, Michigan State University, East Lansing, Michigan 48824, United States.}
\affiliation{Department of Chemical Engineering \& Materials Science, Michigan State University, East Lansing, Michigan 48824, United States.}

\date{\today}

%TC:ignore

\begin{abstract}
Lithium chalcohalide antiperovskites are a promising, non-toxic alternative to lead halide perovskites, with potential as solid electrolytes for Li-ion batteries. we computationally investigate a relatively unexplored anion-interchange mechanism by which cubic Li$_{3}$$AB$ derivatives are obtained from the parent cubic antiperovskite Li$_{3}$$BA$ ($A$ = O, S, Se, Te, Po; $B$ = F, Cl, Br, I). The calculated relative energy landscape provides a useful guide for anion-site selectivity and its role in structural stability. The energetic stability results reveal that the smaller anion inside the octahedron stabilizes the structures. The lattice-dynamic calculations confirm that Li$_{3}$F$A$ ($A$ = Te, Po) and Li$_{3}$O$B$ ($B$ = Cl, Br, I), which are the most energetically stable compounds, are dynamically stable cubic phases without imaginary phonon modes. However, Li$_{3}$FS and Li$_{3}$FSe, while energetically stable, are dynamically unstable at equilibrium and become dynamically stable under triaxial compressive strain. In addition, we report the electronic structure and density of states (DOS) of all compounds, which show a substantial change in band gap upon anion interchange. The strain engineering of the lithium chalcohalide family illustrates how a few percent of the strain can tune the electronic band gap within the electrochemical stability window for solid battery applications. This study unveils essential characteristics of the anion site-interchange mechanism and provides a foundation for the understanding and design of lithium chalcohalide antiperovskites.
\end{abstract}

\maketitle

%TC:endignore

 % \onecolumngrid

\section{Introduction}

Current scientific research on solar energy technologies increasingly emphasizes the development of advanced materials to address global demands for energy efficiency and environmental sustainability. Among renewable energy technologies, solar cells remain one of the most effective systems to directly convert sunlight into electrical energy~\cite{zhang2025exploring, zheng2021antiperovskite, lai2017anti}. In recent years, perovskite-based materials have attracted considerable scientific attention because of their remarkable power-conversion efficiencies and highly tunable physical properties compared to many conventional photovoltaic materials~\cite{zhang2020halide, xia2022antiperovskite, dutra2023computational}. Within this broader field, antiperovskite compounds have emerged as promising candidates for next-generation solar energy applications due to their distinctive structural, electronic, and optical characteristics~\cite{al2024dft, oyeniyi2022electronic}. Antiperovskites generally adopt the chemical formula $X_3AB$, where $X$ represents a cation and $A$ and $B$ denote anions. However, unlike conventional perovskites, antiperovskites exhibit an inversion of ionic positions within the crystal lattice. This structural modification creates unique bonding environments and gives rise to physical properties that are not typically observed in standard perovskite systems~\cite{wang2020antiperovskites, hoffmann2022superconductivity}. Recent progress in first-principles computational methods has allowed researchers to investigate these materials in more detail, facilitating predictions of their suitability for applications in photovoltaics, superconductivity, and energy storage technologies~\cite{uddin2024theoretical, behera2022structural, xiong2024regarding}. Furthermore, experimental studies on alkali metal–based antiperovskites have demonstrated high ionic conductivity along with favorable mechanical stability, highlighting their potential for use in solid-state electrolytes, batteries and advanced electronic devices~\cite{gao2021hydride, gao2023boosting}. Because of their diverse physical properties and multifunctional capabilities, antiperovskite materials continue to attract significant research interest across a wide range of scientific and technological fields. 

For such a purpose in laboratory experiments, Zhao \textit{et al.}~\cite{zhao2012superionic} introduced a novel class of solid electrolytes that feature three-dimensional lithium-ion conduction pathways based on lithium-rich anti-perovskites (LiRAPs). These materials exhibit high ionic conductivity ($\sigma > 10^{-3}~\mathrm{S/cm}$) at room temperature, accompanied by a low activation energy in the range of $0.2$--$0.3~\mathrm{eV}$. As the temperature approaches the melting point, the ionic conductivity increases further, reaching superionic conductivity levels exceeding $\sigma > 10^{-2}~\mathrm{S/cm}$. Moreover, these crystalline antiperovskite materials can be systematically customized through chemical, electronic, and structural modifications to enhance the ionic transport properties. Due to their exceptional Li$^{+}$ conductivity and tunable characteristics, LiRAPs represent promising candidates for high-performance solid electrolytes in advanced electrochemical applications.

\onecolumngrid

\begin{figure}[H]
    \centering
    \includegraphics[width=\textwidth]{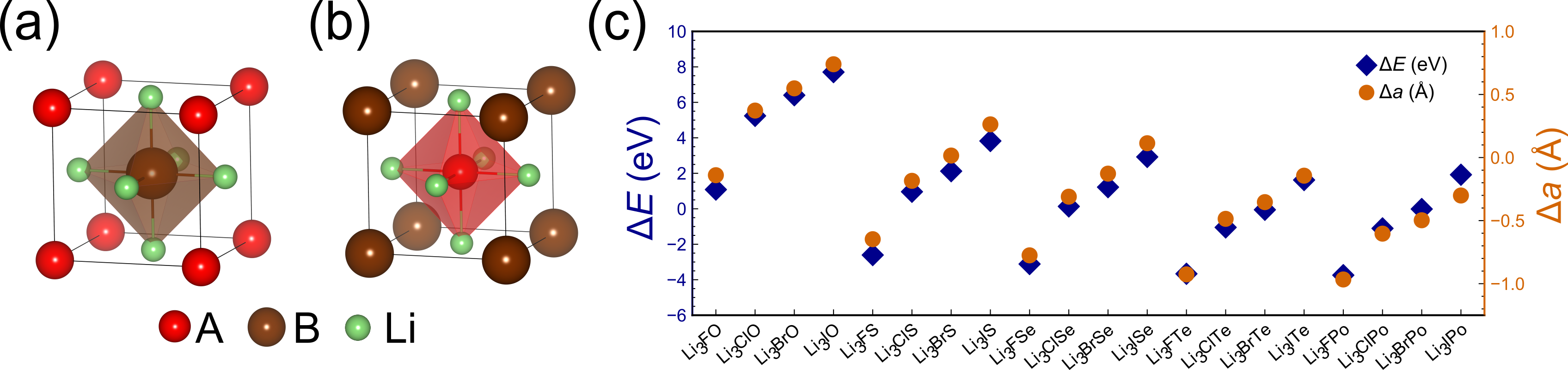}
\caption{
    The crystal structures of cubic \textbf{(a)} antiperovskite (AP); Li$_{3}$$BA$, 
    \textbf{(b)} anion-interchange antiperovskite (AAP); Li$_{3}$$AB$ ($A$ = O$^{2-}$, S$^{2-}$, Se$^{2-}$, Te$^{2-}$, Po$^{2-}$;\ $B$ = F$^{-}$, Cl$^{-}$, Br$^{-}$, I$^{-}$). The AAP approach includes interchanging the A-site $(0,0,0)$ with B-site $(0.5,0.5,0.5)$.
    \textbf{(c)} Variation of the ground state energy difference, $\Delta E = E_{\mathrm{tot}}(\text{AP}) - E_{\mathrm{tot}}(\text{AAP})$, and the lattice constant difference, $\Delta a = a_{AP} - a_{AAP}$, 
    across all compositions. The negative $\Delta E$ and positive $\Delta E$ indicate stable AP and stable AAP, respectively, while negative $\Delta a$ and positive $\Delta a$ indicate larger AAP lattice parameter and larger AP lattice parameter, respectively. 
}
    \label{fig:fig1}
\end{figure}

\twocolumngrid 

In another study, Kim \textit{et al.}~\cite{kim2022exploring} used both theoretical and experimental approaches to investigate the synthesizability of several marginally stable antiperovskite (AP) compounds predicted to exhibit high ionic mobility for Li$^{+}$, Na$^{+}$ and K$^{+}$ ions. Density functional theory (DFT) calculations, combined with the quasi-harmonic approximation, were deployed to evaluate the change in free-energy, $\Delta G_r(T)$, associated with synthesis reactions for 36 alkali-metal-based AP compounds of the form $X_3AZ$ ($X =$ Li, Na or K; $A =$ O, S or Se; $Z =$ F, Cl, Br or I). A linear relationship was identified between the extent of the lattice distortion and the stabilization temperature at which $\Delta G_r(T) = 0$. Their results indicate that AP compounds predicted to exhibit the highest ionic mobility generally require elevated synthesis temperatures to achieve thermodynamic stability.

Ni \textit{et al.}~\cite{ni2024first} used DFT to investigate the structural, electronic, and electrochemical properties of Li$_7$O$_2$Br$_3$. The compound was found to be dynamically stable in its ground state and behaves as a wide-bandgap insulator with a bandgap of approximately 5.83~eV. Compared with Li$_3$OBr, Li$_7$O$_2$Br$_3$ exhibits improved mechanical flexibility, which is advantageous for processing and practical applications. Importantly, it shows a lower migration barrier for Li$^{+}$ (0.26~eV versus 0.40~eV in Li$_3$OBr), mainly due to softened lattice vibrations of Li atoms in the edge layers. Defect engineering analysis further indicates that LiBr-type defects significantly enhance Li$^{+}$ transport. In addition, a pressure--temperature-Gibbs free energy phase diagram was constructed to identify favorable synthesis conditions. Their results highlight Li$_7$O$_2$Br$_3$ as a promising solid electrolyte candidate for next-generation solid-state lithium-ion batteries.

In this work, we present a comprehensive DFT investigation of a series of lithium chalcohalide antiperovskites (APs), Li$_{3}$$BA$, in which halogen anions occupy the octahedral-center site $B$ and chalcogen anions sit at the corner site $A$, together with their anion-interchanged antiperovskites (AAPs) Li$_{3}$$AB$, in which the two anion sites are swapped. To the best of our knowledge, the AP and AAP arrangements have not previously been compared on an equal footing, so it has been unclear which anion prefers which site and how that choice affects their stability and properties. Our central aim is to determine the best anion arrangement and, more broadly, which compositions are energetically and dynamically stable and, therefore, potentially synthesizable. The relative energy, $\Delta E = E_{\mathrm{tot}}(\text{AP}) - E_{\mathrm{tot}}(\text{AAP})$,
quantifies the energy difference between the two arrangements and reveals the optimal atomic configuration, in which the smaller anion prefers to sit at the octahedral center, while the larger anion sits at the corner. The lattice-vibration calculations within the harmonic approximation reveal interesting features governing the structural stability, including the phonon modes associated with the octahedral rotation of BLi$_{6}$. Strain engineering is one potential strategy to stabilize the octahedral rotational mode and to tune the electronic band gap and its character within the electrochemical window to achieve high ionic conductivity. Throughout this work, AP materials are plotted in brown, and AAP materials are plotted in red.

\section{Computational Methods}
All calculations were carried out using the CRYSTAL23 code~\cite{erba2022crystal23} within the density functional theory framework to study the structural, electronic and dynamic properties of cubic lithium-chalcohalide antiperovskites and their AAP variants. Our calculations used the hybrid HSE06~\cite{krukau2006influence} exchange-correlation functional and all-electron Gaussian--type basis functions. Regarding atomic basis sets, triple-zeta valence with polarization basis sets (TZVP), \texttt{pob\_TZVP\_rev2}~\cite{vilela2019bsse}, was used for Li, O, F, S, Cl, Se and Br. For heavier elements, including Te, I and Po, consistent Stuttgart small-core relativistic pseudopotentials (\texttt{pob\_TZVP\_rev2\_SOC}~\cite{laun2022bsse,desmarais2020spin}) were employed.

For full geometry optimizations (atoms and unit cell), the Coulomb and HF exchange series were truncated using tight tolerance factors (TOLINTEG parameters) of 8, 8, 8, 10, 34, with tighter values of 9, 9, 9, 11, 38 for subsequent single point energy and property calculations. The convergence thresholds for total energy were set to $2.72 \times 10^{-7}$~eV and $2.72 \times 10^{-8}$~eV for geometry optimization and single-point energy calculations, respectively. Structural convergence was monitored on the basis of the root-mean-square (RMS) and maximum components of the energy gradients and displacements.

The exchange-correlation contribution was evaluated by numerical integration of the unit cell volume through an extra-extra-large grid (XXGRID), consisting of 99 radial and 1454 angular points in the regions relevant for chemical bonding~\cite{dovesi2023crystal23}. Self-consistent (SCF) convergence was accelerated by the direct inversion of the iterative subspace (DIIS) method~\cite{pulay1980convergence, pulay1982improved}. Brillouin zone integrations were performed using Monkhorst–Pack $k$-point meshes of $24 \times 24 \times 24$ and $30 \times 30 \times 30$ for geometry optimization and single point calculations, respectively, to ensure precision in reciprocal space integration. The electronic band path follows $\Gamma$--X--M--$\Gamma$--R--X\,$\vert$\,R--M high-symmetry points in the Brillouin zone.

Phonon dispersion calculations were carried out using the ﬁnite-displacement method as implemented in CRYSTAL23 within the harmonic approximation~\cite{pascale2004calculation}. Due to the sensitivity of the vibrational frequencies to structural accuracies, rigorous optimization criteria were set as follows: RMS forces of $1.54 \times 10^{-3}$~eV\,\AA $^{-1}$, maximum forces of $2.31 \times 10^{-3}$~eV\,\AA $^{-1}$, RMS displacement of $6.35 \times 10^{-5}$~\AA, and maximum displacement of $9.53 \times 10^{-5}$~\AA. To study the effect of supercell size, phonon dispersion calculations for Li$_3$FS were initially carried out using $2 \times 2 \times 2$ (40 atoms), $3 \times 3 \times 3$ (135 atoms) and $4 \times 4 \times 4$ (320 atoms) supercells. The results did not show significant differences, especially between $3 \times 3 \times 3$ and $4 \times 4 \times 4$ supercells (see Figure~\ref{fig:Benchmark_Phonon}). Therefore, all subsequent phonon calculations in this work were performed using the $3 \times 3 \times 3$ supercells. The visualizations of the atomic structures were generated using the VESTA software package~\cite{momma2011vesta}.

Next, we study the effects of triaxial deformation on these materials. The applied strain percentage $\eta$ is expressed in terms of the 
equilibrium lattice parameter $a_0$ and the strained lattice parameter $a$ as
\begin{equation}
    \eta = \frac{a - a_0}{a_0} \times 100\%,
    \label{eq:strain}
\end{equation}
where $a = b = c \neq a_0$ upon application of the strain. As such, the negative and positive values of $\eta$ in Eq.~(\ref{eq:strain}) correspond to compressive and tensile strain, respectively.

\onecolumngrid

\begin{table}[h!]
\centering
\caption{Calculated electronic and structural properties of the Li$_{3}$$BA$ (AP) and Li$_{3}$$AB$ (AAP) compounds, with cubic lattice parameter ($a_0=b_0=c_0$), unit-cell volume ($V_0$), density ($\rho$), energy band gap ($E_g$), and the relative energy $\Delta E = E_{\mathrm{tot}}(\text{AP}) - E_{\mathrm{tot}}(\text{AAP})$. The parentheses (i) and (d) denote direct and indirect gap, respectively.}
\label{tab:Tableall}
\setlength{\tabcolsep}{4pt}
\resizebox{0.95\textwidth}{!}{%
\begin{tabular}{lcrcc lcrccc}
\toprule
\multicolumn{5}{c}{\textbf{Li$_{3}$$BA$ (AP)}} & \multicolumn{5}{c}{\textbf{Li$_{3}$$AB$ (AAP)}} \\
% \cline{2-5} \cline{7-10}
\cmidrule{2-5} \cmidrule{7-10}
Compound & $a_0$/\AA & $V_0$/\AA$^3$ & $\rho$/g$\cdot$cm$^{-3}$ & $E_g$/eV &
Compound & $a_0$/\AA & $V_0$/\AA$^3$ & $\rho$/g$\cdot$cm$^{-3}$ & $E_g$/eV & $\Delta E$/eV\\
\midrule
Li$_3$FO & 3.524 & 43.772 & 2.126 & 5.530(i) & Li$_3$OF & 3.664 & 49.197 & 1.892 & 6.629(i) & 1.083 \\
Li$_3$ClO & 4.251  & 76.818 & 1.557 & 2.664(i) & Li$_3$OCl & 3.879 & 58.350 & 2.049 & 6.503(i) & 5.239 \\
Li$_3$BrO & 4.517 & 92.136 & 2.090 & 2.165(i) & Li$_3$OBr & 3.968 & 62.480 & 3.082 & 6.157(d) & 6.402 \\
Li$_3$IO & 4.888 & 116.754 & 2.332 & 1.500(d) & Li$_3$OI & 4.148 & 71.366 & 3.815 & 5.484(d) & 7.705 \\
Li$_3$FS & 3.872 & 58.073 & 2.059 & 5.641(i) & Li$_3$SF & 4.519 & 92.303 & 1.296 & 5.472(i) & -2.607 \\
Li$_3$ClS & 4.432 & 87.047 & 1.679 & 3.825(i) & Li$_3$SCl & 4.616 & 98.385 & 1.485 & 5.269(i) & 0.961 \\
Li$_3$BrS & 4.664  & 101.487 & 2.159 & 3.320(i) & Li$_3$SBr & 4.648 & 100.431 & 2.181 & 5.038(d) & 2.121 \\
Li$_3$IS & 5.008 & 125.577 & 2.379 & 2.578(d) & Li$_3$SI & 4.745 & 106.828 & 2.797 & 4.604(d) & 3.822 \\
Li$_3$FSe & 3.977 & 62.897 & 3.167 & 5.162(d) & Li$_3$SeF & 4.752 & 107.320 & 1.856 & 5.326(i) & -3.114 \\
Li$_3$ClSe & 4.500 & 91.095 & 2.478 & 3.808(i) & Li$_3$SeCl & 4.810 & 111.271 & 2.029 & 5.119(i) & 0.134 \\
Li$_3$BrSe & 4.712 & 104.604 & 2.856 & 3.303(d) & Li$_3$SeBr & 4.839 & 113.335 & 2.636 & 4.835(d) & 1.220 \\
Li$_3$ISe & \num{5.02971} & 127.241 & 2.974 & \num{2.6165}(d) & Li$_3$SeI & 4.916 & 118.820 & 3.185 & 4.424(d) & 2.924 \\
Li$_3$FTe & \num{4.18624454} & 73.362 & 3.847 & \num{3.6704}(d) & Li$_3$TeF & 5.112 & 133.597 & 2.112 & 4.864(d) & -3.664 \\
Li$_3$ClTe & \num{4.62243954} & 98.767 & 3.126 & \num{3.6744}(d) & Li$_3$TeCl & 5.108 & 133.281 & 2.316 & 4.747(d) & -1.050 \\
Li$_3$BrTe & \num{4.79691342} & 110.378 & 3.458 & \num{3.1349}(d) & Li$_3$TeBr & 5.150 & 136.630 & 2.794 & 4.497(d) & -0.055 \\
Li$_3$ITe & \num{5.06721027} &130.108 & 3.546 & \num{2.5581}(d)& Li$_3$TeI & 5.211 & 141.509 & 3.261 & 4.145(d) & 1.622 \\
Li$_3$FPo & \num{4.22084310} & 75.197 & 5.499 & \num{3.0969}(d) & Li$_3$PoF & 5.187 & 139.588 & 2.962 & 4.732(d) & -3.743 \\
Li$_3$ClPo & \num{4.57197090} & 95.567 & 4.605 &\num{2.8187}(i) & Li$_3$PoCl & 5.175 & 138.621 & 3.174 & 4.555(d) & -1.106 \\
Li$_3$BrPo & \num{4.72160478} & 105.261 & 4.874 & \num{2.3653}(d) & Li$_3$PoBr & 5.219 & 142.137 & 3.609 & 4.320(d) & -0.014 \\
Li$_3$IPo & \num{4.96909895} & 122.696 & 4.831 & \num{1.8044}(d) & Li$_3$PoI & 5.270 & 146.363& 4.050 & 3.952(d) & 1.920 \\
\bottomrule
\end{tabular}
}  
\end{table}

\twocolumngrid

\section{Results and Discussions}
\paragraph{Crystal structure and energetic stability:}
The cubic lithium-chalcohalide Li$_{3}$$BA$ antiperovskite (AP) structure (space group $Pm\bar{3}m$, no.~221), shown in Figure~\ref{fig:fig1}(a), consists of the $B$ anion (halide, $\mathrm{B}^{-} \in \{\mathrm{F}^{-},\,\mathrm{Cl}^{-},\,
\mathrm{Br}^{-},\,\mathrm{I}^{-}\}$) at the body-center site, coordinated by six Li$^{+}$ ions located at the face-center positions, while the $A$ anion (chalcogen, $\mathrm{A}^{2-} \in \{\mathrm{O}^{2-},\,\mathrm{S}^{2-},\,\mathrm{Se}^{2-},\,\mathrm{Te}^{2-},\,\mathrm{Po}^{2-}\}$) occupies the corner site with 12-fold-coordination. Figure~\ref{fig:fig1}(b) illustrates the anion-interchange antiperovskite (AAP) variant, Li$_{3}$$AB$, in which the chalcogen ($A$) and halogen ($B$) sites are swapped between the corner and body-centered positions, modifying the chemical bonding environment and the electronic structure, including changes in the size and behavior of the band-gap, such as the direct-to-indirect transition and its reverse. 

This site-interchange approach doubles the number of computationally predicted compounds, with 20 Li$_{3}$$BA$ (AP) and 20 Li$_{3}$$AB$ (AAP) structures. Such a site-interchange approach raises important questions related to which anion prefers which lattice site and the influence of this choice on structural stability and electronic band structure. Our simulations answer these questions by generating quantitative estimates of the relative energies of the AP and AAP phases. Overall, our results provide a useful systematic guide for anion-site selectivity and its role in structural stability.

First, we benchmarked our calculated lattice parameters compared to available experimental data. The optimized lattice parameters of Li$_{3}$OCl and Li$_{3}$OBr are 3.88~\text{\AA} and 3.97~\text{\AA}, respectively, in excellent agreement with the experimental values of 3.91~\text{\AA} and 4.02~\text{\AA} reported previously by Zhao \textit{et~al.}~\cite{zhao2012superionic}. This experimental atomic arrangement corresponds to our AAP convention, in which O$^{2-}$ occupies the octahedral center $B$ site and Cl$^{-}$ or Br$^{-}$ occupies the corner $A$ site. However, this arrangement does not always hold for other chalcogens. Therefore, a close analysis is required to identify which atomic arrangement is energetically and dynamically stable. The complete DFT results of all 40 compositions are presented in Table~\ref{tab:Tableall}. 

\onecolumngrid

\begin{figure}[H]
    \centering
    \includegraphics[width=0.8\textwidth]{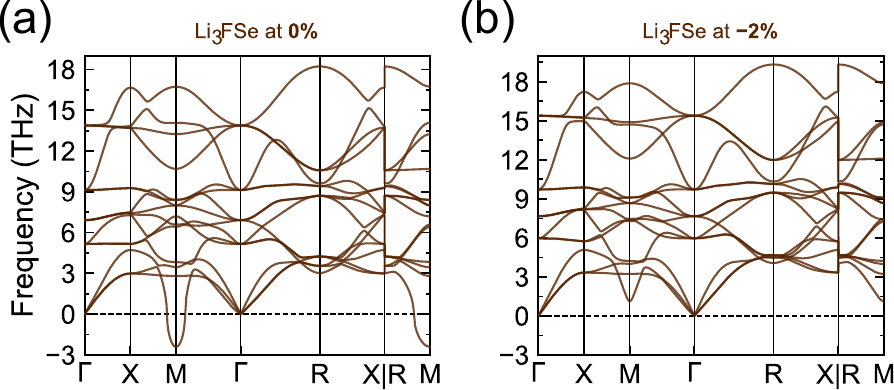}
    \caption{Phonon dispersion curves of Li$_3$FSe under \textbf{(a)} 0\% strain and \textbf{(b)} $-2\%$ strain, computed using a $3\times3\times3$ supercell. Negative frequency on the vertical axis represents an imaginary mode.}
    \label{fig:Phononselect}
\end{figure}

\twocolumngrid

Second, to investigate the energetic stability of the structures after interchanging the anions, we performed hybrid HSE06 DFT calculations on each pair of Li$_{3}$$BA$ (AP) and Li$_{3}$$AB$ (AAP) and compared their total energies and lattice parameters, as shown in Figure~\ref{fig:fig1}(c). The relative energy, ranging from $-3.74$ to $+7.71$~eV, is evaluated using $\Delta E = E_{\mathrm{tot}}(\text{AP}) - E_{\mathrm{tot}}(\text{AAP})$ where  negative $\Delta E$ indicates that the AP structure is more stable, and positive $\Delta E$ is equivalent to the stable AAP structure. As shown in Figure~\ref{fig:fig1}(c), the energy difference increases along the halogen series to produce many stable AAP topologies, while it decreases down the chalcogen series to shift some stable compositions toward the AP topology. As a consequence, we found that 8 compounds favor the AP topology (Li$_{3}$FS, Li$_{3}$FSe, Li$_{3}$FTe, Li$_{3}$ClTe, Li$_{3}$BrTe, Li$_{3}$FPo, Li$_{3}$ClPo, and Li$_{3}$BrPo), while 12 compounds favored the AAP topology (Li$_{3}$OF, Li$_{3}$OCl, Li$_{3}$OBr, Li$_{3}$OI, Li$_{3}$SCl, Li$_{3}$SBr, Li$_{3}$SI, Li$_{3}$SeCl, Li$_{3}$SeBr, Li$_{3}$SeI, Li$_{3}$TeI, and Li$_{3}$PoI). To better understand the physical origin of the trend of energy variation in Figure~\ref{fig:fig1}(c), the following observations are viable; (i) The largest energy differences are observed for oxyhalide structures, where O$^{2-}$ at the body-centered site is paired with Cl$^{-}$, Br$^{-}$, or I$^{-}$ at the corner site, with the magnitude increasing in that order. Hence, all oxyhalide structures are energetically stable in the AAP phase with oxygen at the octahedral center site, and any halogen can occupy the corner site. Similarly, all iodine-based compounds are energetically stable in the AAP phase with I$^{-}$ in the corner and any chalcogen at the center site. (ii) For chalcogen-based structures other than O$^{2-}$, the smallest energy differences are observed when F$^{-}$ occupies the octahedral center site and is paired with S$^{2-}$, Se$^{2-}$, Te$^{2-}$, or Po$^{2-}$ at the corner site, making them an energetically stable AP topology. (iii) For chlorine and bromine-based compounds, intermediate energy differences are observed with energetic stability depending on the paired chalcogen other than O$^{2-}$; such that the lighter chalcogen (S$^{2-}$ or Se$^{2-}$) favors AAP and the heavier chalcogen (Te$^{2-}$ or Po$^{2-}$) favors the AP phase. These points clearly show that the structural stability is determined not only by placing the divalent chalcogen at the body-center, as in oxyhalide compounds, but also depends on other factors such as anion size and site occupancy.

In general, the trend of the computed energy difference, Figure~\ref{fig:fig1}(c), shows a clear site preference controlled by the anion size, such that the smaller anion is always preferred at the body-centered site, while the larger anion is favored at the corner position. The only deviation from this trend is observed for Li$_{3}$SeCl and Li$_{3}$PoI, which behave in the opposite way, with the smaller anion in the corner and the larger anion in the center of the body. These structures are regarded as metastable cubic phases due to the anomalous expansion of the lattice site by the oversize anion. In general, the origin of the energetic stability can be primarily attributed to electrostatic interactions; placing the smaller anion at the octahedral body-centered site corresponds to the shorter anion–cation distance, favoring Coulombic attraction between the central anion and surrounding Li$^{+}$ ions and hence enhancing stability. To further explore the consequence of the phenomenon of anion interchange on the lattice parameters, we calculated the relative lattice parameter [Figure~\ref{fig:fig1}(c)], $\Delta a = a_{AP} - a_{AAP}$, which follows the same pattern as the relative total energy but with values of $\Delta a$ ranging from $-0.97$ to $+0.74$~\text{\AA}. Negative values indicate that the AAP phase has larger lattice parameters, while positive values indicate that the AP topology has larger lattice parameters. This result reveals a consistent trend such that placing the larger anion at the body-centered site expands the lattice parameters, except in some energetically competitive compounds where anions have relatively close atomic numbers (e.g., Li$_{3}$OF and Li$_{3}$SCl). The above analysis identifies which AP or AAP structure has the lowest energy in the cubic phase; however, phonon calculations are needed to test for dynamical stability. It is also important to note that energetically unstable cubic structures can be stabilized by doping, strain, or raising the temperature, so it is useful to know the cubic
phase properties for this purpose. It is also useful to understand the energetics and phonon properties of the parent symmetric structure when considering various instabilities a composition may have. Therefore, we now analyze phonon calculations to
understand the dynamical stability of AP and AAP systems.

\onecolumngrid

\begin{figure}[H]
    \centering
    \includegraphics[width=0.8\textwidth]{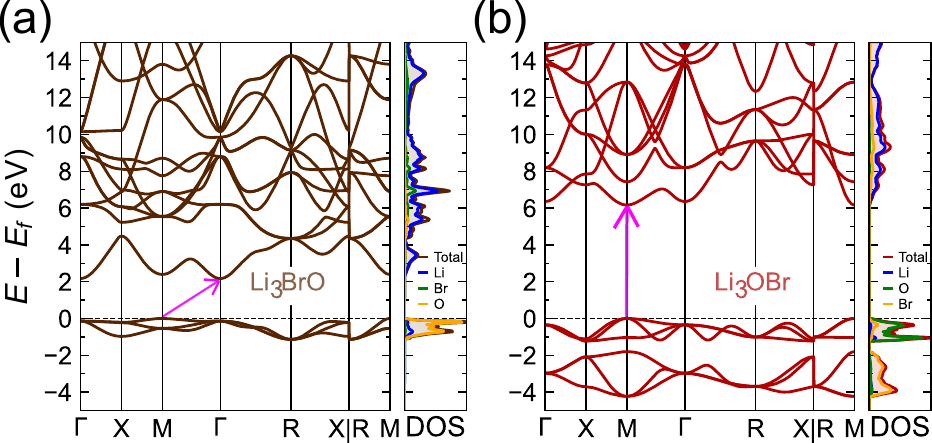}
    \caption{Electronic band structures and density of states (DOS) for \textbf{(a)} \ce{Li_3BrO} (brown) and \textbf{(b)} \ce{Li3OBr} (red), showing that anion swapping leads to change in the band gap. The valence band maximum is set to 0~eV.}
    \label{fig:selectedbands}
\end{figure}

\twocolumngrid

\paragraph{Phonon band structure:}
Research attempts have increased tremendously on both the theoretical~\cite{dutra2023computational,ni2024first,fang2017li,luo2024theoretical,kim2019correlating,kim2025lattice,zhang2013ab} and experimental~\cite{zhao2012superionic,fujii2021alkali,lai2017anti,gao2023boosting,kim2022exploring,emly2013phase,zhang2014high,milan2025lithium,yin2020synthesis} sides to validate lithium-based APs, although mainly for oxyhalides. In contrast, many lithium chalcohalide APs remain relatively unexplored.

To investigate and validate their dynamical stability, we obtained phonon dispersion spectra along high-symmetry directions in the Brillouin zone for all compounds considered in this work. The results are presented in Figure~\ref{fig:Phonon_AP} for all AP structures and in Figure~\ref{fig:Phonon_AAP} for their corresponding AAP phases. Of the 20 AP compositions, only the fluorine-based members Li$_{3}$FTe, and Li$_{3}$FPo exhibit phonon spectra with no imaginary frequencies, indicating that they are lattice-dynamically stable in the cubic antiperovskite phase, with the acoustic branches dominating up to about 3--4~THz. Their optical and acoustic branches give similar frequency characteristics across the high-symmetry points. In contrast, Li$_{3}$FS and Li$_{3}$FSe are qualitatively different, such that their atomic displacements show imaginary modes (lattice instabilities) at the high-symmetry points M and R, associated with FLi$_{6}$ octahedra instabilities, which does not disrupt the overall crystal lattice~\cite{gao2021hydride, cancellieri2016polaronic}. These imaginary frequencies may be resolved by including anharmonic phonon effects at finite temperature~\cite{tadano2015self,chen2015anharmonicity} or by applying external pressure up to about 5 GPa, as in the case of Li$_{3}$HS and Na$_{3}$FSe reported by Gao \textit{et al}~\cite{gao2021hydride} and Fujii \textit{et al}~\cite{fujii2021alkali}, respectively. As a result, we conducted a proof-of-concept trial to stabilize them by applying triaxial compressive strain, and their results are discussed below. Therefore, Li$_{3}$FS and Li$_{3}$FSe antiperovskites could be synthesized in principle but most likely in phases with lower energy tilted distortions. The remaining AP structures show significant imaginary frequencies in their phonon dispersion and are therefore dynamically unstable (see  Table~\ref{tab:Tablefrq}). 

For AAP structures, the oxyhalide antiperovskites Li$_{3}$OCl, Li$_{3}$OBr, and Li$_{3}$OI exhibit fully positive phonon frequencies, with acoustic branches extending to about 3--5~THz, confirming their dynamical stability. Their optical branches display similar frequency distributions across the high-symmetry points. However, the acoustic branches are distributed progressively in the low-frequency region with increasing halide atomic size from Cl$^{-}$ to Br$^{-}$ to I$^{-}$. This difference is due to the effect of atomic weight. All other AAP structures, including Li$_{3}$OF, show significant imaginary frequencies and are also dynamically unstable. For all dynamically unstable AP and AAP structures apart from Li$_{3}$FS and Li$_{3}$FSe, all high symmetry points show imaginary modes, with the most negative frequencies occurring at M~$(\frac{1}{2},\frac{1}{2},0)$, ranging from $-11.88$ to $-6.07$~THz (See  Table~\ref{tab:Tablefrq}).

Together, the dynamically stable members in both the AP and the AAP families agree with the most energetically stable compositions identified at both extremes of Figure~\ref{fig:fig1}(c) which are the highest and lowest relative energies, whereas the most unstable modes are found in compounds with intermediate relative energies. According to Table~\ref{tab:Tablefrq}, imaginary frequencies everywhere in the Brillouin zone mean there is an atomic displacement direction that further lowers the energy, leading to instability against small perturbations. Figures~\ref{fig:Phononselect}(a) and (b) present the phonon band structures of a representative compound, Li$_{3}$FSe, at 0\% and $-2\%$ triaxial strain. At a 0\% strain, Li$_{3}$FSe is energetically stable, with $\Delta E = -3.11$~eV relative to Li$_{3}$SeF, but the phonon mode at the M~$(\frac{1}{2},\frac{1}{2},0)$ point exhibits an imaginary frequency of $-2.41$~THz, indicating a lower-symmetry structure arising from the rotational mode of the FLi$_{6}$ octahedra. However, the imaginary frequency disappears under compressive strain (at $-2\%$), stabilizing the cubic structure. The structures of the phonon band for other strain tests, ranging from $-5\%$ to $+5\%$ are presented in  Figures~\ref{fig:PhononstrainLi3FSe}. Similarly, Li$_{3}$FS (with 0\% deformation) exhibits imaginary phonon frequencies of $-4.56$~THz and $-2.28$~THz at the M and R points, respectively. However, applying compressive strain stabilizes the structure, as shown in Figures~\ref{fig:PhononstrainLi3FS}.

\onecolumngrid

\begin{figure}[H]
    \centering
    \includegraphics[width=0.8\textwidth]{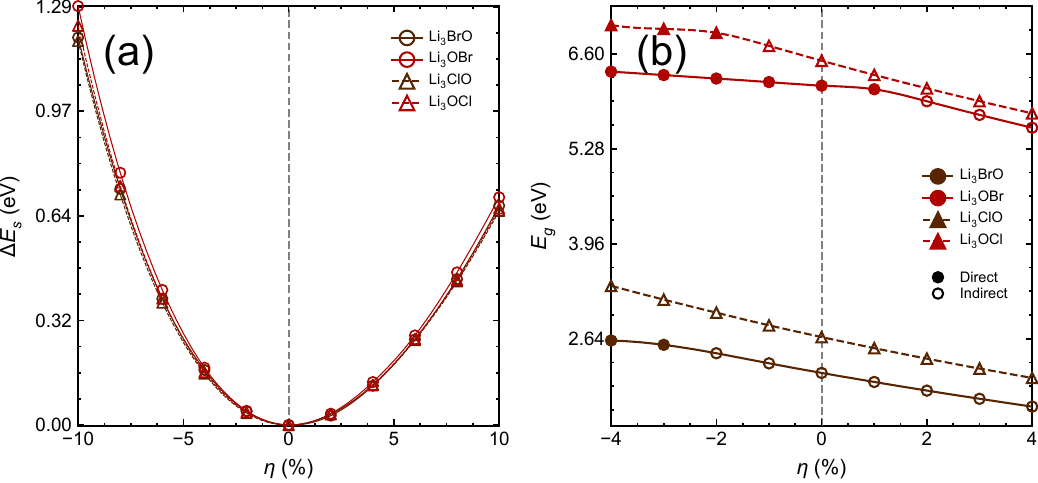}
    \caption{\textbf{(a)} Relative strain energy, $\Delta E_{s} = E_{\mathrm{strained}} - E_{\mathrm{equilibrium}}$, as a function of applied strain for Li$_3$BrO, Li$_3$OBr, Li$_3$ClO, and Li$_3$OCl. \textbf{(b)} Strain-induced variation of the band gap energy $E_g$ for the same compounds.}
    \label{fig:RelativeStrainEnergy}
\end{figure}

\twocolumngrid

\paragraph{Electronic property and strain effect:}
Here, we summarize the electronic band structures of all predicted compositions, as presented in Table~\ref{tab:Tableall}. In general, there is a substantial change in the band gap after the anions are interchanged compared to that of the parent AP structures. This change relates to the site occupancy of the anions, which
modifies the degree of orbital overlap. For example, Figures~\ref{fig:selectedbands}(a) and (b) show the electronic band structure and DOS of representative compounds containing O and Br ions. Figure~\ref{fig:selectedbands}(a) displays the band structure and DOS of Li$_{3}$BrO in the AP phase, where O$^{2-}$ sits at the corner site and Br$^{-}$ at the octahedral center. Here, the valence band maximum (VBM) is formed by O~p-orbitals and the conduction band minimum (CBM) by Li~p-orbitals, with an indirect band gap of 2.17~eV. Swapping the anion sites (Figure~\ref{fig:selectedbands}(b)) shifts the Br~p-orbitals upward to combine with the O~p-orbitals to form the VBM, with the dominant contribution coming from the O~p-orbitals, while the highest contribution to the CBM remains solely the Li p-orbitals, producing a direct band gap of 6.16~eV. The interchange of the anions and the associated orbital shifts lead to a change in lattice geometry and band gap size and type. The electronic band structures and DOS of the remaining structures are shown in  Figure~\ref{fig:stablebands} (energetically stable members) and  Figure~\ref{fig:unstablebands} (energetically unstable members). A generic feature in this class of cubic lithium chalcohalide antiperovskite is that for AP materials, the chalcogenide p-orbitals dominate the VBM, while for AAP cases there is a mixing of p-orbitals of both anions (see Table~\ref{tab:Tableorbital}).

An important property of solid electrolyte materials is that they must be electronically insulating and ionically conducting. Therefore, the band gap of a solid electrolyte material must be at least as large as the electrochemical stability window, which determines the voltage range of the battery. In practice, a band gap larger than about 5~eV is sufficient for the large electrochemical stability window~\cite{emly2013phase,xiao2019computational}.  Aside from Li$_{3}$O$A$ ($A$ = halogen), that exhibit significant ionic conductivity character ($E_g > 5$~eV) in their AAP phase (see Table~\ref{tab:Tableall}), the following compounds also show relatively large bandgaps above 5~eV: 5.64~eV for Li$_{3}$FS, 5.27~eV for Li$_{3}$SCl, 5.04~eV for Li$_{3}$SBr, 5.16~eV for Li$_{3}$FSe and 5.12~eV for Li$_{3}$SeCl. These wide bandgaps suggest good electronic insulating properties to meet the requirements as a solid electrolyte. With this in mind, understanding the impact of the strain on their electronic bandgap and band type is important.

Initially, we calculated the electronic band structures of the equilibrium Li$_{3}$OCl and Li$_{3}$OBr in both the AP and AAP phases. As shown in Table~\ref{tab:Tableall},  Figure~\ref{fig:stablebands} and  Figure~\ref{fig:unstablebands}, Li$_{3}$OCl shows an indirect band gap E$_g$ of 6.50~eV (AAP phase),  with the VBM and CBM at M and \(\Gamma\) points, respectively. In contrast, the VBM and CBM  of Li$_{3}$OBr are located at point M of the Brillouin zone, with a direct band gap E$_g$ of 6.16~eV (AAP phase). Our calculated band gaps for Li$_{3}$OCl and Li$_{3}$OBr are in excellent agreement with previous reports~\cite{emly2013phase,singh2026electrochemical,wu2018bulk,luo2024theoretical}. As shown in  Figure~\ref{fig:SelectedStrainBand}, a slight strain of $-1\%$ applied to Li$_{3}$OCl retains the indirect character (E$_g$ = 6.71~eV), while the strain $-2\%$ changes the character of the band to a direct band gap (E$_g$ = 6.90~eV), with both the CBM and the VBM located at the M point of the Brillouin zone. This shows that compressive strain significantly influences the electronic band structure, especially at the M point. In general, it is well established that a large band gap Li$_{3}$OCl is an excellent ionic conductor with negligible electronic conductivity, but it remains uncertain whether the band gap is direct or indirect, because no experimental evidence is available for this. Our hybrid calculations HSE06 reveal that Li$_{3}$OCl has an indirect band gap (E$_g$ = 6.50~eV) in the stable phase of unstrained AAP and becomes direct (E$_g$ = 6.90~eV) under compressive strain $2\%$. As a result, it is worth exploring the effects of strain in detail; we then expanded the strain analysis from $-10\%$ to $+10\%$ at intervals of $2\%$ for Li$_{3}$OA (A = Cl$^{-}$ and Br$^{-}$), in both the phases AP and AAP (see Figure~\ref{fig:RelativeStrainEnergy} ), using Equation~\ref{eq:strain}. To demonstrate the feasibility of the strain-induced effect within the elastic limit, we compared the relative total strain energies as a function of strain, as shown in Figure~\ref{fig:RelativeStrainEnergy}(a). The energy-strain curves show quadratic behavior, indicating greater mechanical flexibility and an elastic limit between approximately $-4\%$ and $+4\%$ for reversible deformation. 

We next restrict our calculation to the elastic regime to examine the direct-to-indirect band-type transition and its reverse. As shown in Figure~\ref{fig:RelativeStrainEnergy}(b), the band gap decreases with tensile strain and increases with compressive strain. Within this regime, Li$_{3}$BrO undergoes an indirect-to-direct transition at $3\%$ compressive strain, while its AAP counterpart, Li$_{3}$OBr, exhibits a direct-to-indirect transition at $2\%$ tensile strain. In contrast, Li$_{3}$ClO does not show a band-type transition in the strain range considered, while its AAP phase, Li$_{3}$OCl, undergoes an indirect-to-direct transition at $2\%$.

We then investigated whether this behavior generalizes across the other compounds, and we applied the same elastic-regime strain to all energetically stable structures. The corresponding strain-dependent band gap variations are summarized in 
Figures~\ref{fig:strainbandsall}(a) -- (d). Across all systems, the band gap varies linearly with strain, with only a small deviation at the strain value where a band-type transition occurs. This is consistent with Figure~\ref{fig:RelativeStrainEnergy}(b), where compressive (tensile) strain induces an indirect-to-direct (direct-to-indirect) transition. Most compounds exhibit a decreasing
band gap with increasing strain, except in some structures, such as the iodine-based compounds
Li$_{3}$$A$I ($A$ = S, Se, Te, Po), which increase with strain instead. Furthermore, all iodine-based structures retain a direct band gap throughout the strain range, with no transition observed at any strain value considered. The deviation from linearity that marks a band-type transition originates from strain-induced changes in the local bonding and orbital environment near the band edges. 

The lithium chalcohalide structures predicted in this work could be optimized further for phase stability and ionic-conductivity performance through several routes already demonstrated in halide (anti)perovskites, including site/chemical doping~\cite{fujii2021alkali,huang2024insight,gao2023boosting,liang2025multiple,qian2026sulfur}, engineering of AP/AAP interfaces~\cite{singh2026electrochemical,zhang2023bilayer,deng2022bilayer}, mixed-anion substitution~\cite{zhao2012superionic, fang2017li} and defect engineering~\cite{stegmaier2017li+,mouta2016li+,wang2023managing}. Combined with the anion-interchange mechanism and strain tuning established in this work, such structural and chemical design strategies give a clear path to achieve the full potential of these materials as next-generation solid-state electrolytes.

\section{Conclusions}
Using hybrid DFT, we investigated the energetic stability, lattice-dynamical stability, electronic structure, and strain engineering of lithium chalcohalide antiperovskites, including their corresponding anion-interchange variants. The main findings are summarized as follows.

Energetic stability of AP relative to AAP structures identified some stability preference: 
(i) all oxyhalide structures show high energy differences and are energetically stable in AAP phase with oxygen at the octahedral center. (ii) all fluorine-based structures paired with chalcogens other than O$^{2-}$ show smaller energy differences and are stable in the AP phase, with F$^{-}$ at the site of the octahedral center. (iii) intermediate energy differences are observed for Cl$^{-}$ -- and Br$^{-}$ -- based compounds paired with chalcogens other than O$^{2-}$, with stability depending on the paired chalcogen. Overall, the energetic stability confirms that the stability depends not solely on the positioning of the divalent chalcogen inside the octahedron, as in the oxyhalide class, but also on anion size and site occupancy. Generally, placing the larger anion at the lattice corner site and the smaller anion inside the octahedron stabilizes the structures, resulting in 20 most energetically stable structures. 

The phonon dispersion calculations further confirm that only a subset of compositions (Li$_{3}$FTe, Li$_{3}$FPo, Li$_{3}$OCl, Li$_{3}$OBr and Li$_{3}$OI) are dynamically stable in cubic phase, in agreement with the most energetically stable structures obtained from the relative total energy calculations. In addition, Li$_{3}$FSe and Li$_{3}$FS are found to be dynamically stable in the cubic phase under triaxial compressive strain.

The electronic band structures and density of states show a consistent pattern in which the anion p-states govern the top of the valence bands, while Li p-states dominate the bottom of the conduction bands across all compositions. Several materials, including oxyhalides, sulfur halides, and some Se-based structures, display insulating band gaps greater than 5~eV, making them promising wide-band-gap solid electrolytes whose properties can be engineered via controlled lattice distortions, anion-site ordering, and octahedral rotational distortions. Triaxial strain is shown to significantly modify the electronic band gap across the lithium chalcohalide family, with most compounds exhibiting a decreasing band gap with increasing strain, except in some structures, such as the iodine-based compounds. Within the elastic regime, several compounds further exhibit a reversible indirect-to-direct band-type transition and its inverse, highlighting the triaxial strain as an effective route for tuning both the magnitude and type of the band gap in these antiperovskite materials.

% \section{REFERENCES}

\bibliography{references}

\clearpage
\newpage

\renewcommand{\thepage}{S\arabic{page}}
\renewcommand{\thesection}{S\arabic{section}}
\renewcommand{\thetable}{S\arabic{table}}
\renewcommand{\thefigure}{S\arabic{figure}}
\renewcommand{\theequation}{S\arabic{equation}}

\setcounter{page}{1}
\setcounter{section}{0}
\setcounter{figure}{0}
\setcounter{table}{0}
\setcounter{equation}{0}

\renewcommand{\theHfigure}{S\arabic{figure}}
\renewcommand{\theHtable}{S\arabic{table}}
\renewcommand{\theHequation}{S\arabic{equation}}
\renewcommand{\theHsection}{S\arabic{section}}

\onecolumngrid

\begin{center}
{\LARGE \textbf{Supporting Information (SI)}}
\end{center}

% \vspace{1em}
\noindent\rule{\textwidth}{1.5pt}

\begin{center}
\Large\textbf{Can Strain or Anion Interchange Make an Unstable Structure Stable? Energetics, Lattice Dynamics and Strain-Tunable Band Gaps of Lithium Chalcohalide Antiperovskites (Li$_{3}$$BA$) and their Anion Interchange Variants (Li$_{3}$$AB$)}

\vspace{1em}
\large
Ismail A. Buliyaminu$^{a}$, Ehsan Gowdini$^{a}$, Phillip Duxbury$^{a}$ and Jose L. Mendoza-Cortes$^{a,b,*}$

\vspace{1em}
\normalsize
$^{a}$~Department of Physics and Astronomy,
Michigan State University, East Lansing, MI 48824, USA\\[0.3em]
$^{b}$~Department of Chemical Engineering and Materials Science,
Michigan State University, East Lansing, MI 48824, USA\\[0.5em]
$^{*}$~E-mail: jmendoza@msu.edu
\end{center}

\noindent\rule{\textwidth}{1.5pt}

\tableofcontents
% \newpage
% \vfill

\phantomsection
\addcontentsline{toc}{section}{Supporting Information}

\vspace{1em}

\begin{figure}[H]
    \centering
    \includegraphics[width=\columnwidth]{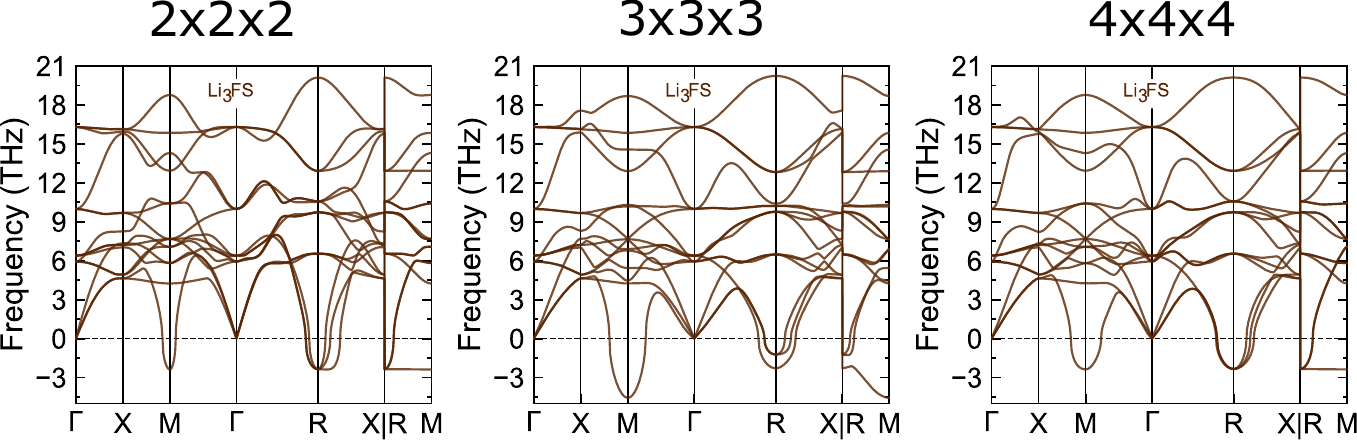}
\caption{Phonon dispersion curves of Li$_3$FS, computed using a $2\times2\times2$ (40 atoms), $3\times3\times3$ (135 atoms), and $4\times4\times4$ (320 atoms) supercells.}
    \label{fig:Benchmark_Phonon}
\end{figure}

\begin{figure}[H]
    \centering
    \includegraphics[width=\columnwidth]{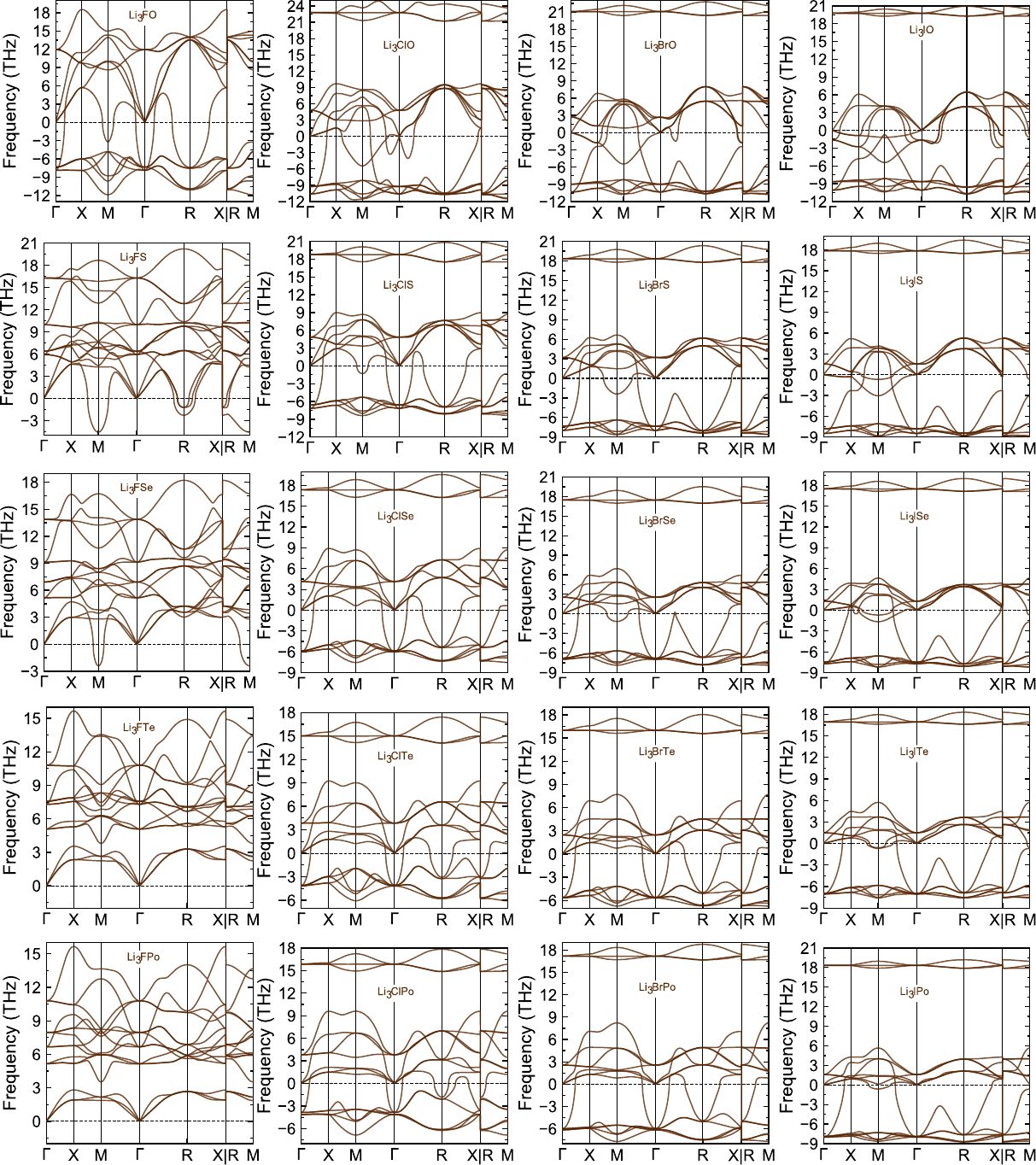}
\caption{Phonon dispersion curves of AP compounds, computed using a $3\times3\times3$ supercell.}
    \label{fig:Phonon_AP}
\end{figure}

\begin{figure}[H]
    \centering
    \includegraphics[width=\columnwidth]{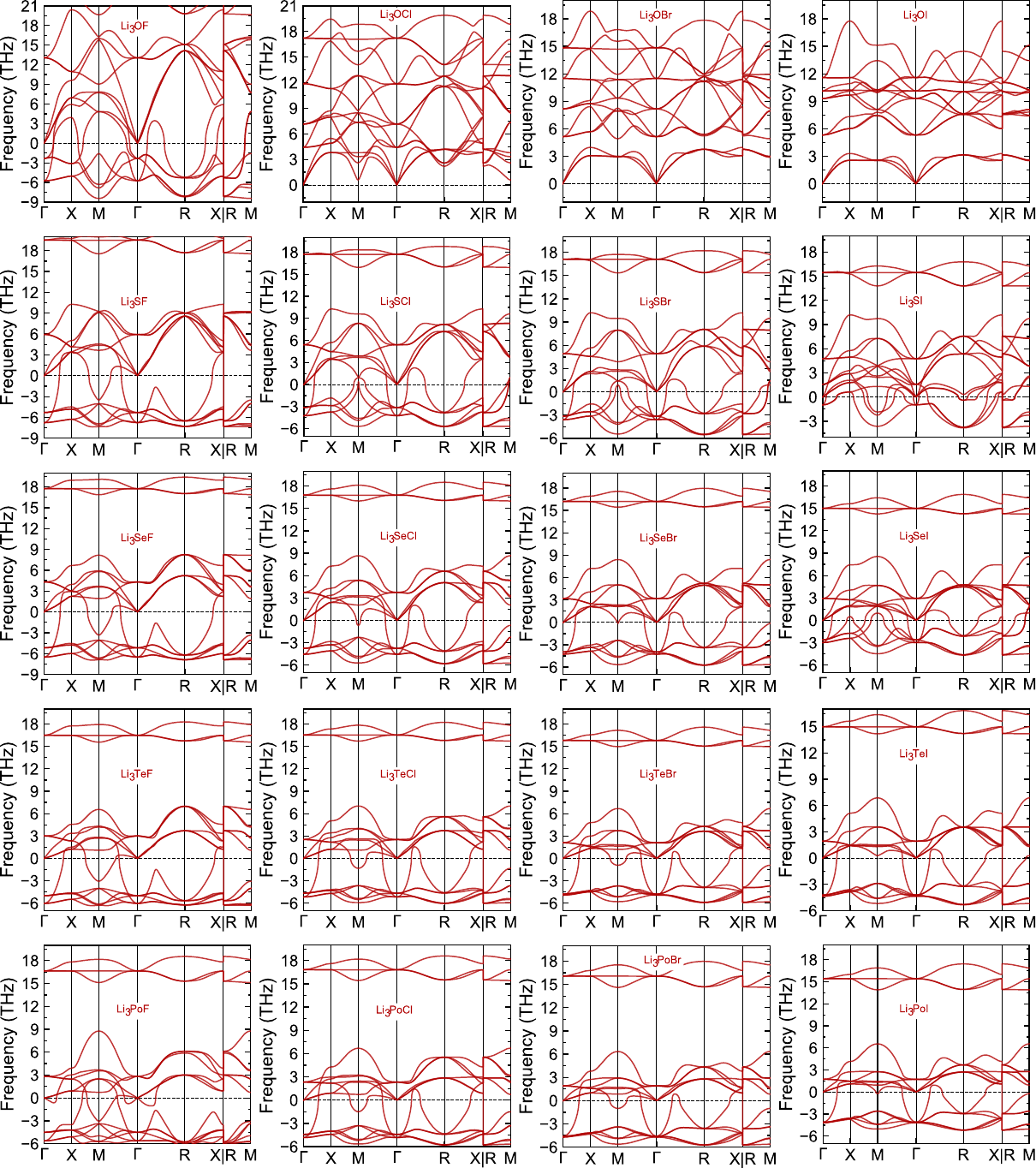}
\caption{Phonon dispersion curves of AAP compounds, computed using a $3\times3\times3$ supercell.}
    \label{fig:Phonon_AAP}
\end{figure}

\begin{figure}[H]
    \centering
    \includegraphics[width=\columnwidth]{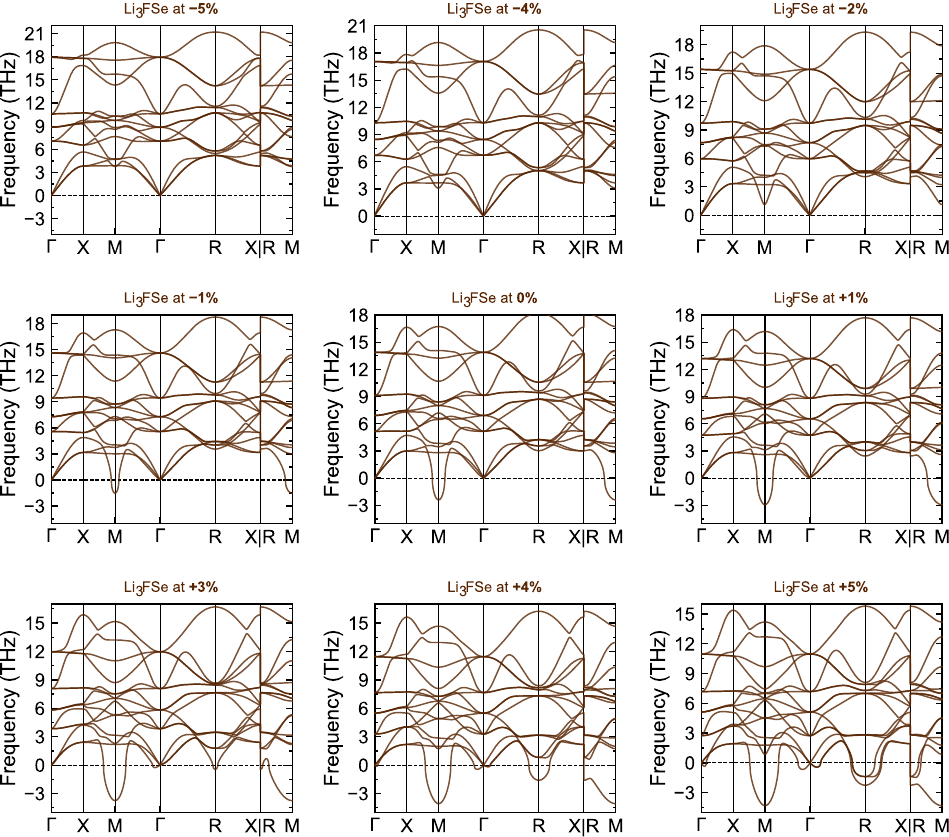}
    \caption{Phonon dispersion curves of Li$_3$FSe under $-5\%$ to $+5\%$ strain, computed using a $3\times3\times3$ supercell. Negative frequency on the vertical axis represents an imaginary mode}
    \label{fig:PhononstrainLi3FSe}
\end{figure}

\begin{figure}[H]
    \centering
    \includegraphics[width=\columnwidth]{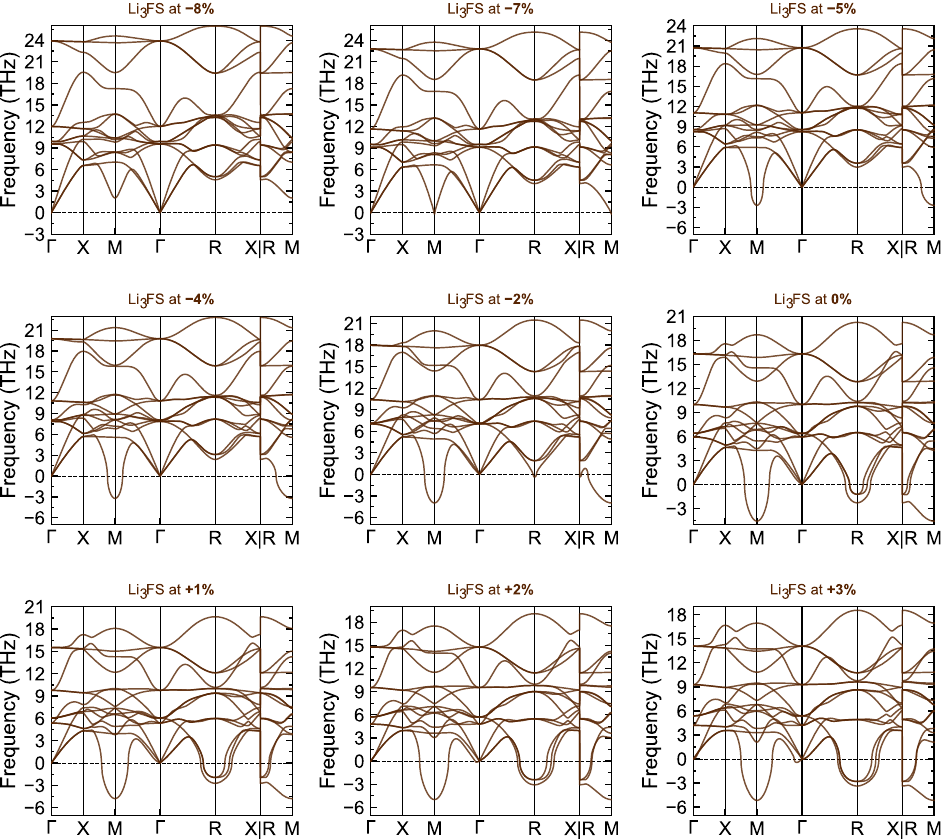}
    \caption{Phonon dispersion curves of Li$_3$FS under -8\% to +3\% strain, computed using a $3\times3\times3$ supercell. Negative frequency on the vertical axis represents an imaginary mode}
    \label{fig:PhononstrainLi3FS}
\end{figure}

\begin{figure}[H]
    \centering
    \includegraphics[width=\textwidth]{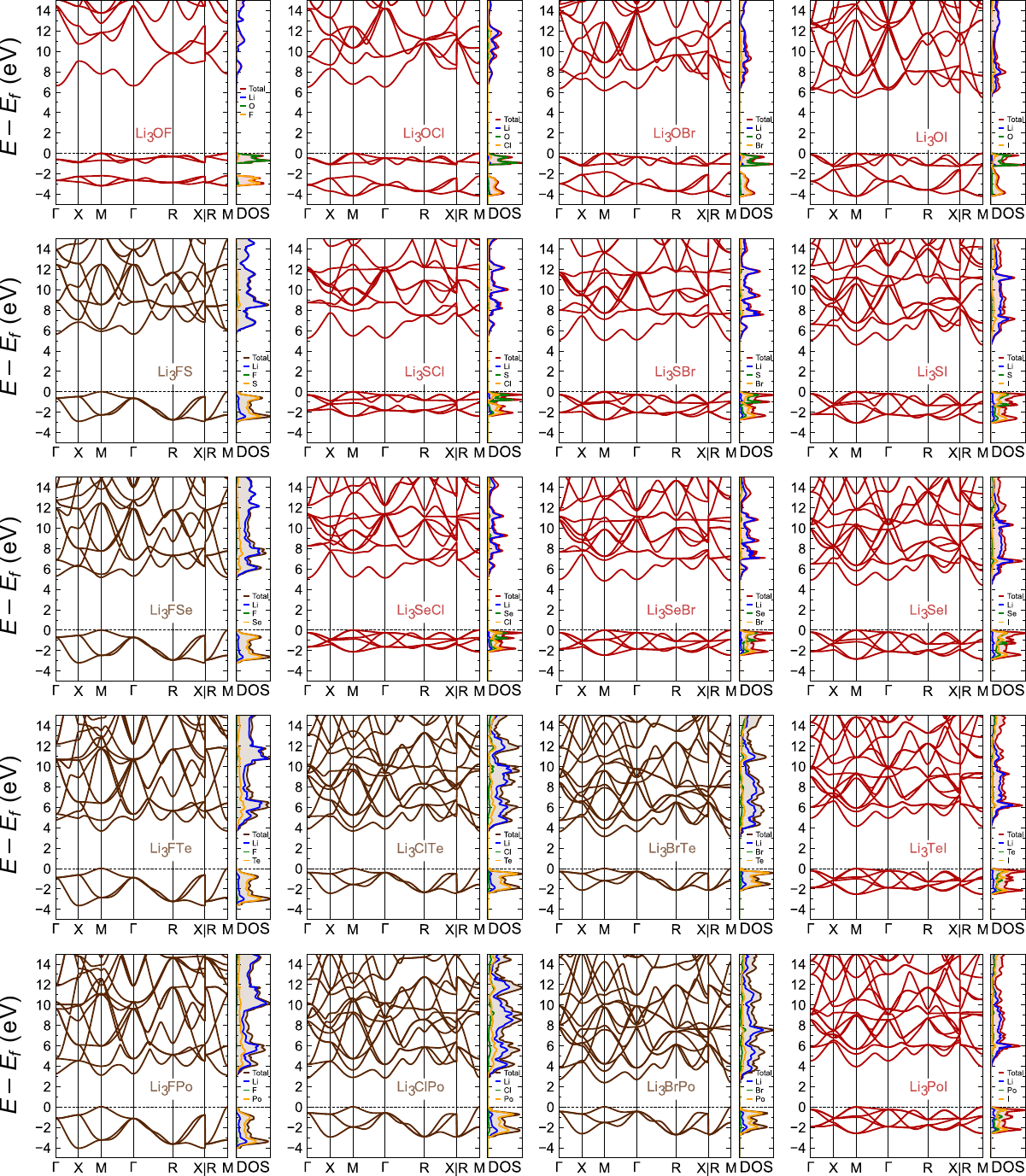}
    \caption{Electronic band structures and density of states (DOS) of the energetically stable structures, Li$_{3}$$BA$ (brown) and Li$_{3}$$AB$ (red). The valence band maximum is set to zero eV.}
    \label{fig:stablebands}
\end{figure}

\begin{figure}[H]
    \centering
    \includegraphics[width=\textwidth]{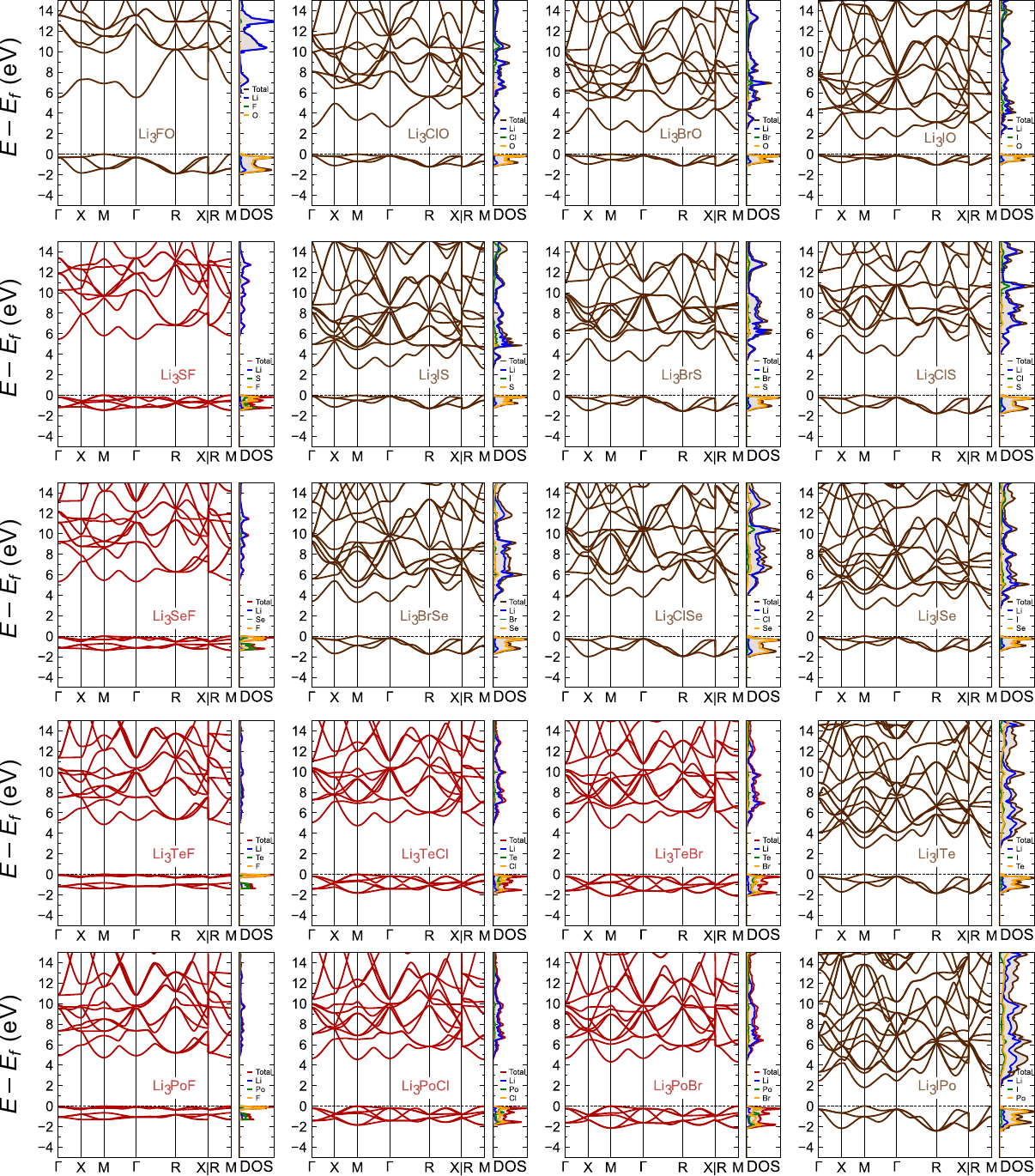}
    \caption{Electronic band structures and density of states (DOS) of the energetically unstable structures, $L_3BA$ (brown) and $L_3AB$ (red). The valence band maximum is set to zero eV.}
    \label{fig:unstablebands}
\end{figure}

\begin{table}[h!]
\centering
\caption{Minimum phonon frequencies (THz) at high-symmetry \textbf{k}-points of the Brillouin zone and total DFT/HSE06 energy for all structures in the cubic $Pm\bar{3}m$ phase, calculated using a $3\times3\times3$ supercell. Negative frequencies indicate imaginary phonon modes, signifying dynamic instability of the cubic phase. All symmetry points with imaginary modes have the most negative frequency at M~$(\frac{1}{2},\frac{1}{2},0)$, consistent with a zone-boundary octahedral-tilting instability. The total energy corresponds to the converged SCF ground-state energy per unit cell.}
\label{tab:Tablefrq}
\renewcommand{\arraystretch}{1.2}
\setlength{\tabcolsep}{4pt}
\resizebox{\textwidth}{!}{%
\begin{tabular}{lcccc|lccccc}
\hline
\multicolumn{5}{c|}{Li$_{3}$$BA$--frequencies (THz)} & \multicolumn{5}{c}{Li$_{3}$$AB$--frequencies (THz)} \\
\cline{2-5} \cline{7-10}
Compound & $\Gamma$ & X & M & R &
Compound & $\Gamma$ & X & M & R & $\Delta E$/eV\\
\hline
Li$_3$FO & $-7.882$ & $-7.883$ & $-11.884$ & $-11.110$ & Li$_3$OF & $-5.748$ & $-5.033$ & $-8.449$ & $-8.209$ & 1.083 \\
Li$_3$ClO & $-10.600$  & $-9.849$ & $-11.159$ & $-10.744$ & Li$_3$OCl & $-0.005$ & $3.805$ & $0.597$ & $2.202$ & 5.239 \\
Li$_3$BrO & $-10.338$ & $-9.691$ & $-10.764$ & $-10.652$ & Li$_3$OBr & $-0.004$ & $3.065$ & $2.895$ & $3.765$ & 6.402 \\
Li$_3$IO & $-10.081$ & $-9.451$ & $-10.387$ & $-10.723$ & Li$_3$OI & $-0.005$ & $2.563$ & $2.515$ & $3.150$ & 7.705 \\
Li$_3$FS & $-0.0051$ & $4.6395$ & $-4.5609$ & $-2.2779$ & Li$_3$SF & $-6.752$ & $-6.271$ & $-7.309$ & $-7.316$ & -2.607 \\
Li$_3$ClS & $-7.160$ & $-6.637$ & $-8.386$ & $-8.133$ & Li$_3$SCl & $-4.246$ & $-3.739$ & $-5.690$ & $-5.741$ & 0.961\\
Li$_3$BrS & $-7.979$  & $-7.495$ & $-8.629$ & $-8.413$ & Li$_3$SBr & $-3.593$ & $-3.122$ & $-5.446$ & $-5.545$ & 2.121 \\
Li$_3$IS & $-8.489$ & $-8.039$ & $-8.836$ & $-8.838$ & Li$_3$SI & $-0.970$ & $0.627$ & $-3.633$ & $-3.836$ & 3.822 \\
Li$_3$FSe & $-0.006$ & $2.9931$ & $-2.4072$ & $3.0232$ & Li$_3$SeF & $-6.501$ & $-5.988$ & $-6.851$ & $-6.906$ & -3.114\\
Li$_3$ClSe & $-6.006$ & $-5.596$ & $-7.542$ & $-7.365$ & Li$_3$SeCl & $-4.545$ & $-4.085$ & $-5.708$ & $-5.813$ & 0.134\\
Li$_3$BrSe & $-6.971$ & $-6.636$ & $-7.878$ & $-7.835$ & Li$_3$SeBr & $-4.253$ & $-3.855$ & $-5.659$ & $-5.756$ & 1.220 \\
Li$_3$ISe & $-7.730$ & $-7.423$ & $-8.196$ & $-8.113$ & Li$_3$SeI & $-2.971$ & $-2.405$ & $-4.473$ & $-4.665$ & 2.924 \\
Li$_3$FTe & $-0.007$ & $2.341$ & $2.233$ & $3.298$ & Li$_3$TeF & $-6.031$ & $-5.479$ & $-6.269$ & $-6.260$ & -3.664 \\
Li$_3$ClTe & $-4.168$ & $-3.795$ & $-6.065$ & $-5.821$ & Li$_3$TeCl & $-5.167$ & $-4.686$ & $-5.917$ & $-6.033$ & -1.050 \\
Li$_3$BrTe & $-5.671$ & $-5.364$ & $-6.740$ & $-6.722$ & Li$_3$TeBr & $-4.916$ & $-4.472$ & $-5.801$ & $-5.916$ & -0.055\\
Li$_3$ITe & $-7.049$ & $-6.809$ & $-7.574$ & $-7.579$ & Li$_3$TeI & $-4.357$ & $-3.855$ & $-5.178$ & $-5.338$ & 1.620 \\
Li$_3$FPo & $-0.002$ & $1.915$ & $1.867$ & $2.673$ & Li$_3$PoF & $-5.618$ & $-5.088$ & $-5.757$ & $-6.317$ & -3.743 \\
Li$_3$ClPo & $-4.131$ & $-4.127$ & $-6.851$ & $-6.255$ & Li$_3$PoCl & $-4.832$ & $-4.410$ & $-5.662$ & $-5.788$ & -1.106 \\
Li$_3$BrPo & $-6.140$ & $-5.933$ & $-7.705$ & $-7.527$ & Li$_3$PoBr & $-4.692$ & $-4.337$ & $-5.609$ & $-5.736$ & -0.014 \\
Li$_3$IPo & $-8.015$ & $-7.805$ & $-8.644$ & $-8.750$ & Li$_3$PoI & $-4.177$ & $-3.706$ & $-5.036$ & $-5.204$ & 1.920 \\
\hline
\end{tabular}
}  
\end{table}

\begin{figure}[H]
    \centering
    \includegraphics[width=\columnwidth]{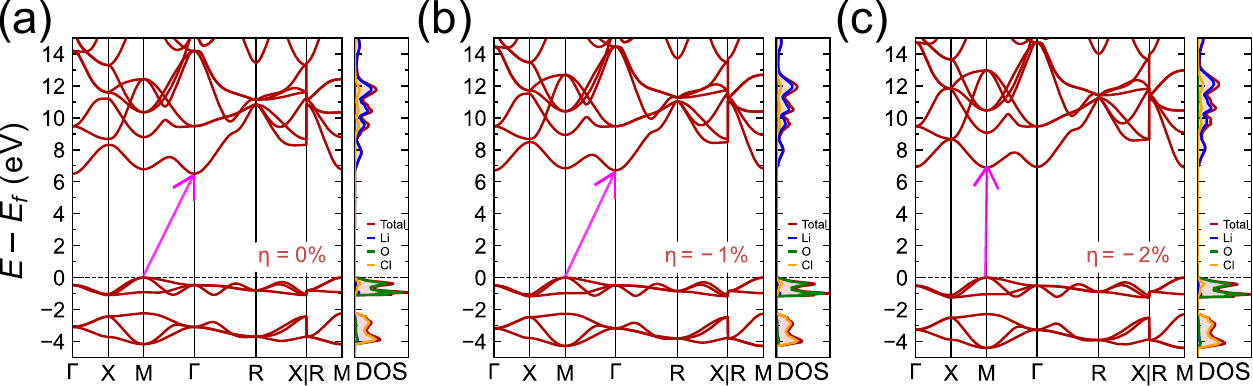}
\caption{Electronic Band structures and density of states (DOS) of \ce{Li_3OCl} under a slight compressive strain ranging from $0\%$ to $-2\%$. The zero energy is set to the VBM}
    \label{fig:SelectedStrainBand}
\end{figure}

\begin{figure}[H]
    \centering
    \includegraphics[width=\textwidth]{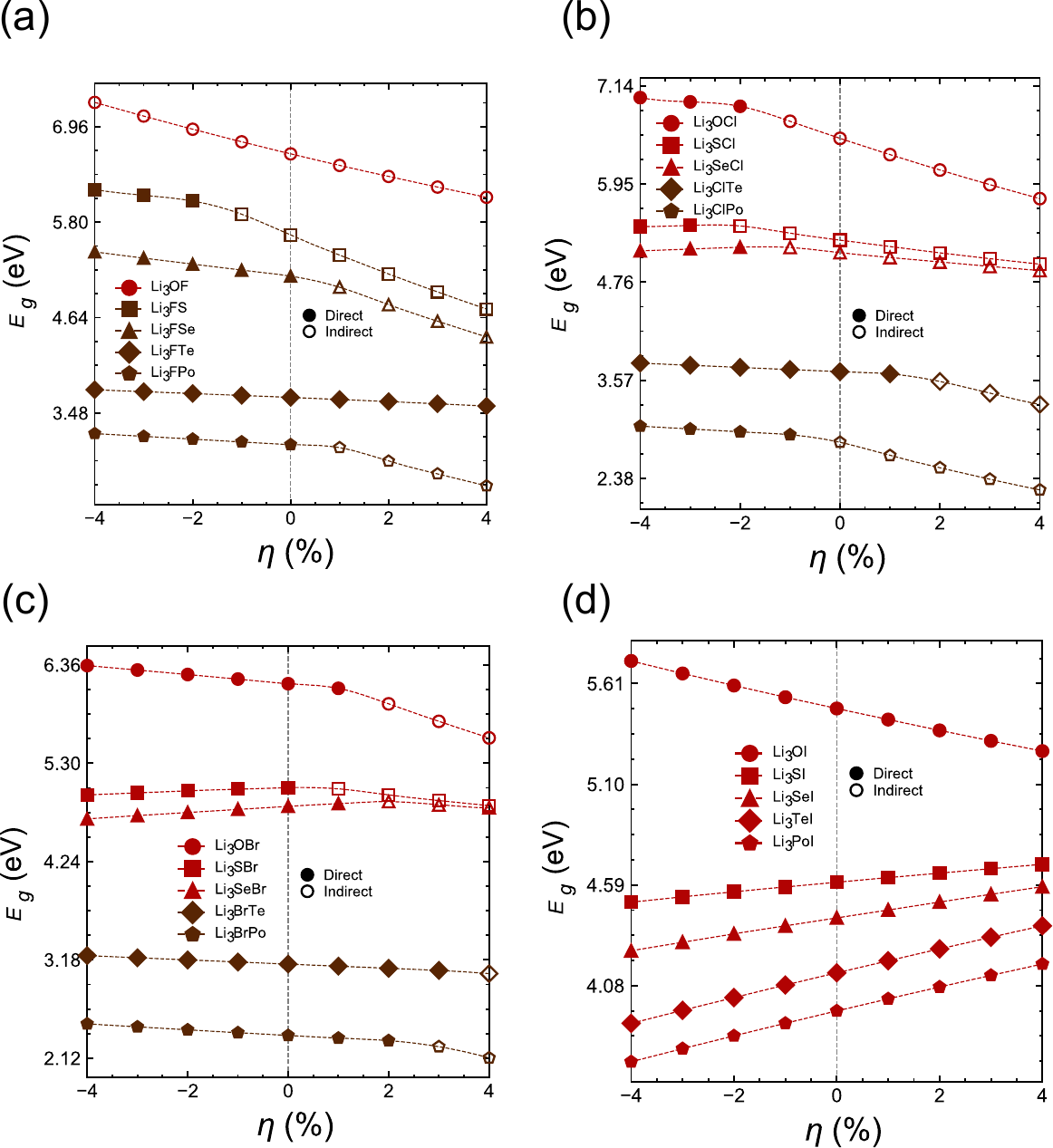}
    \caption{Strain-induced variation of the band gap energy $E_g$ for all energetically stable \textbf{(a)} fluorine-based, \textbf{(b)} chlorine-based, \textbf{(c)} bromine-based, and \textbf{(d)} iodine-based structures.}
    \label{fig:strainbandsall}
\end{figure}

\newpage

\begin{table}[h!]
\centering
\caption{The contribution of atomic orbitals to the valence band (VB) and conduction band (CB) of the  Li$_{3}$$BA$ and Li$_{3}$$AB$ ($A$ = O, S, Se, Te, Po;\ $B$ = F, Cl, Br, I). In the format $X\_y$, where X is the name of the element and y is the atomic orbital, the orbital that contributes the most is highlighted in red for the valence band and in blue for the conduction band.}
\label{tab:Tableorbital}
\begin{tabular}{lll|lll}
\hline
\multicolumn{3}{c|}{Li$_{3}$$BA$} & \multicolumn{3}{c}{Li$_{3}$$AB$} \\
\cline{1-3} \cline{4-6}
Compound & $VB$ & $CB$ & Compound & $VB$ & $CB$ \\
\hline
Li$_3$FO    & \textcolor{red}{$O\_p$}   & \textcolor{blue}{$Li\_p$} & Li$_3$OF    & \textcolor{red}{$O\_p$}, \textcolor{red}{$F\_p$}   & \textcolor{blue}{$Li\_s$} \\
Li$_3$ClO   & \textcolor{red}{$O\_p$}   & \textcolor{blue}{$Li\_s$} & Li$_3$OCl   & \textcolor{red}{$O\_p$}, \textcolor{red}{$Cl\_p$}  & \textcolor{blue}{$Li\_p$} \\
Li$_3$BrO   & \textcolor{red}{$O\_p$}   & \textcolor{blue}{$Li\_p$} & Li$_3$OBr   & \textcolor{red}{$O\_p$}, \textcolor{red}{$Br\_p$}  &  \textcolor{blue}{$Li\_p$} \\
Li$_3$IO    & \textcolor{red}{$O\_p$}   & \textcolor{blue}{$Li\_p$} & Li$_3$OI    & \textcolor{red}{$O\_p$}, \textcolor{red}{$I\_p$}   & \textcolor{blue}{$Li\_p$} \\
Li$_3$FS    & \textcolor{red}{$S\_p$}   & \textcolor{blue}{$Li\_p$} & Li$_3$SF    & \textcolor{red}{$S\_p$}, \textcolor{red}{$F\_p$}   & \textcolor{blue}{$Li\_s$} \\
Li$_3$ClS   & \textcolor{red}{$S\_p$}   & \textcolor{blue}{$Li\_p$} & Li$_3$SCl   & \textcolor{red}{$S\_p$}, \textcolor{red}{$Cl\_p$}  & \textcolor{blue}{$Li\_p$} \\
Li$_3$BrS   & \textcolor{red}{$S\_p$}   & \textcolor{blue}{$Li\_p$} & Li$_3$SBr   & \textcolor{red}{$S\_p$}, \textcolor{red}{$Br\_p$}  & \textcolor{blue}{$Li\_p$} \\
Li$_3$IS    & \textcolor{red}{$S\_p$}   & \textcolor{blue}{$Li\_p$} & Li$_3$SI    & \textcolor{red}{$S\_p$}, \textcolor{red}{$I\_p$}   & \textcolor{blue}{$Li\_p$} \\
Li$_3$FSe   & \textcolor{red}{$Se\_p$}  & \textcolor{blue}{$Li\_p$} & Li$_3$SeF   & \textcolor{red}{$Se\_p$}, \textcolor{red}{$F\_p$}  & \textcolor{blue}{$Li\_s$} \\
Li$_3$ClSe  & \textcolor{red}{$Se\_p$}  & \textcolor{blue}{$Li\_p$} & Li$_3$SeCl  & \textcolor{red}{$Se\_p$}, \textcolor{red}{$Cl\_p$} & \textcolor{blue}{$Li\_p$} \\
Li$_3$BrSe  & \textcolor{red}{$Se\_p$}  & \textcolor{blue}{$Li\_p$} & Li$_3$SeBr  & \textcolor{red}{$Se\_p$}, \textcolor{red}{$Br\_p$} & \textcolor{blue}{$Li\_p$} \\
Li$_3$ISe   & \textcolor{red}{$Se\_p$}  & \textcolor{blue}{$Li\_p$} & Li$_3$SeI   & \textcolor{red}{$Se\_p$}, \textcolor{red}{$I\_p$}  & \textcolor{blue}{$Li\_p$} \\
Li$_3$FTe   & \textcolor{red}{$Te\_p$}  & \textcolor{blue}{$Li\_p$} & Li$_3$TeF   & \textcolor{red}{$Te\_p$}, \textcolor{red}{$F\_p$}  & \textcolor{blue}{$Li\_p$} \\
Li$_3$ClTe  & \textcolor{red}{$Te\_p$}  & \textcolor{blue}{$Li\_p$} & Li$_3$TeCl  & \textcolor{red}{$Te\_p$}, \textcolor{red}{$Cl\_p$} & \textcolor{blue}{$Li\_p$} \\
Li$_3$BrTe  & \textcolor{red}{$Te\_p$}  & \textcolor{blue}{$Li\_p$} & Li$_3$TeBr  & \textcolor{red}{$Te\_p$}, \textcolor{red}{$Br\_p$} & \textcolor{blue}{$Li\_p$} \\
Li$_3$ITe   & \textcolor{red}{$Te\_p$}  & \textcolor{blue}{$Li\_p$} & Li$_3$TeI   & \textcolor{red}{$Te\_p$}, \textcolor{red}{$I\_p$}  & \textcolor{blue}{$Li\_p$} \\
Li$_3$FPo   & \textcolor{red}{$Po\_p$}  & \textcolor{blue}{$Li\_p$} & Li$_3$PoF   & \textcolor{red}{$Po\_p$}, \textcolor{red}{$F\_p$}  & \textcolor{blue}{$Li\_p$} \\
Li$_3$ClPo  & \textcolor{red}{$Po\_p$}  & \textcolor{blue}{$Li\_p$} & Li$_3$PoCl  & \textcolor{red}{$Po\_p$}, \textcolor{red}{$Cl\_p$} & \textcolor{blue}{$Li\_p$} \\
Li$_3$BrPo  & \textcolor{red}{$Po\_p$}  & \textcolor{blue}{$Li\_p$} & Li$_3$PoBr  & \textcolor{red}{$Po\_p$}, \textcolor{red}{$Br\_p$} & \textcolor{blue}{$Li\_p$} \\
Li$_3$IPo   & \textcolor{red}{$Po\_p$}  & \textcolor{blue}{$Li\_p$} & Li$_3$PoI   & \textcolor{red}{$Po\_p$}, \textcolor{red}{$I\_p$}  & \textcolor{blue}{$Li\_p$} \\
\hline
\end{tabular}
\end{table}

\end{document}